\documentclass[twocolumn, showpacs]{revtex4}
\usepackage{graphics}
\usepackage{amsmath}
\usepackage{overpic}
\usepackage{hyperref}
\usepackage{rotating}
\newcommand{\ket}[1]{|#1\rangle}
\newcommand{\bra}[1]{\langle #1|}

\begin{document}
\title{Geometric Phase and Quantum Phase Transition in the Lipkin-Meshkov-Glick model}
\author{H. T. Cui}
\email{cuiht@student.dlut.edu.cn}
\author{K. Li, X. X. Yi} \affiliation{Department of
Physics, Dalian University of Technology, Dalian 116024, China}
\date{\today}
\begin{abstract}
The relation between the geometric phase  and quantum  phase
transition has been discussed in the Lipkin-Meshkov-Glick model. Our
calculation shows the ability of geometric phase of the ground state
to mark quantum phase transition in this model. The possibility of
the geometric phase or its derivatives as the universal order
parameter of characterizing quantum phase transitions has been also
discussed.
\end{abstract}
\maketitle

\section{introduction}
Recently the understanding of quantum phase transition
\cite{sachdev} has emerged from the fundamentals  of quantum
mechanics, especially from the entanglement point of view
\cite{preskill}. The ground state structure have been shown to be
affected appreciably by the critical points, as illustrated
initially in the one-dimensional Ising model with transverse
magnetic field \cite{osterloh}. Intriguing by these pioneering
works, a great deal of efforts have attributed to this region
\cite{wu,vidal,gu}. It originates from the belief that quantum phase
transition should be connected to the entanglement or  its
derivatives in many-body systems \cite{preskill}. However the
correspondence between the entanglement and quantum phase transition
in many-body systems is ambiguous \cite{yang, vidal1}, that is
because of the absence of the proper measurement of entanglement in
many-body systems. On the other hand, geometric phase \cite{berry}
as a measurement of the curvature of the Hilbert space, has been
first connected with the quantum phase transition in the
one-dimension spin-chain system \cite{carollo}, in which the
topological character of the relative geometric phase between the
first excited state and the ground state have been shown the ability
of detecting the critical points. Furthermore, Zhu extended this
study under the thermodynamic limit and found that the geometric
phase of the ground state in the one-dimensional XY model was
non-analytical and its derivative with the coupling constant was
divergent closed to critical points \cite{zhu}. Moreover the scaling
behavior of the geometric phase of ground state clearly
distinguished two different types of quantum phase transitions at
the critical point in the one-dimensional XY model. Consequently the
general theory about the relation between geometric phase of the
ground state and the critical point has been constructed
\cite{hamma}, in which the topological property of Berry's loops
including critical points in the parameter space has been shown the
ability of detecting the critical points.

Although the theory is rapidly developing, the verifications of
examples only focus on the one-dimension XY spin-chain systems. The
reason is that XY model, which have been shown first-order phase
transition closed to the critical point, can be converted into
spinless fermionic  system by Jordan-Wigner transformation, and the
energy spectrum can be determined exactly \cite{lieb}. However the
interacting between spins in this model is short-range (the
neighbor-nearest coupling) and the anisotropy of Heisenberg
interaction is indispensable for the construction of long-range
order \cite{lieb}. Recently the Berry phase in Dick model has been
examined under thermodynamics limit \cite{plastina}. The authors
showed that the Berry phase displayed the non-analyticity close to
critical point where this model exhibits a second-order phase
transition, and its derivative with the coupling constant was also
discontinued and showed a cusp closed to this point. The
discontinuity of Berry phase is very similar to that in XY model,
but we should point out that the two model belong to different
orders of quantum phase transitions respectively. In this paper, we
will show a special situation, in which the geometric phase of the
ground state  behaves differently and itself distinguishes the
different quantum phase transitions.

In particular we study the geometric phase in a system of spins with
a collective coupling described by the Lipkin-Meshkov-Glick (LMG)
model \cite{lmg}, which first introduced forty years ago in nuclear
physics. This model has received much attention because of his
apparent simplicity and popularity in the past. It provides a simple
description of the tunneling of bosons between two degenerate levels
and can thus be used to describe many physical systems such as
two-mode Bose-Einstein condensates \cite{cirac} or Josephson
junctions \cite{josephson}. More recently the entanglement in this
model has received great attention because of available numerical
calculations and plentiful  phase diagram \cite{vidal2, vidal1}.
Under the thermodynamics limit, its phase diagram can be simply
established by a semiclassical approach \cite{botet}. For large
particle number N but finite, the situation is complicated and the
numerical analysis was implemented using the continuous unitary
transformations \cite{dusuel}. The significant difference between
this model and XY model is the long-range interaction, and the
system cannot be converted into the spinless fermionic system. Hence
it is of great interest to study the relation of geometric phase and
phase transition in this model.  As will display in the remaining of
this paper, the geometric phase of ground state in this model
behaves differently and reflects faithfully  the existence of the
critical points.

The paper is organized as follows. In Sec. II, we first introduce
the LMG model, and in the limit of large N but finite, we obtain the
ground state analytically by the Holstein-Primakoff representation
and calculate the geometric phase. In order to show the university
of our results, in Sec. III, we study the phase transition in a more
complex situation. It is surprising that the geometric phase of
ground state detects very rigorously the critical points. Finally we
discuss the implications of our results and the differences from the
XY model.

\section{the Lipkin-Meshkov-Glick model: biaxial case}
The LMG model describes a set of N spins half coupled to all others
with a strength independent of the position and the nature of the
elements and a  magnetic field in the $z$ direction. The Hamiltonian
can be written
\begin{equation}\label{h1}
H= - \frac{1}{N}(S^2_x + \gamma S^2_y) - h S_z,
\end{equation}
in which $S_{\alpha}=\sum_{i=1}^{N}\sigma^i_{\alpha}/2 (\alpha=x, y,
z)$ and the $\sigma_{\alpha}$ is the Pauli operator, N is the total
particle number in this system. The prefactor $1/N$ is essential to
ensure the convergence of the free energy per spin in the
thermodynamic limit. For any anisotropy parameter $\gamma \in[0,1]$,
the Hamiltonian \eqref{h1} preserves the total spin and does not
couple the state having spin pointing in the direction perpendicular
to the field, namely
\begin{equation}
[H, \textbf{S}^2]=0,   [H, \prod_{i=1}^{N}\sigma_z^i]=0.
\end{equation}
An important character that differentiates LMG model from the XY
model is the long-rang interaction between particles, which induces
a second-order phase transition at $h=1$ when $N\rightarrow\infty$
\cite{botet}.

The diagonalization  of Eq. \eqref{h1} can be obtained by
introducing the Holstein-Primakoff representation of the spin
operator and then truncate the resulting bosonic Hamiltonian to
lowest order \cite{dusuel}. Consequently we diagonalize it thanks to
the Bogoliubov transformation. The first thing is to perform a
rotation of the spin operators around the $y$ direction, that makes
the $z$ axis along the so-called semiclassical magnetization
\cite{dusuel} in which the Hamiltonian Eq. \eqref{h1} has the
minimal value in the semiclassical approximation. This can be done
as
\begin{equation}
\left( \begin{array}{c} S_x \\S_y \\S_z
\end{array} \right)= \left(
\begin{array}{ccc} \cos\theta & 0 & -\sin\theta \\ 0 & 1 & 0 \\
\sin\theta & 0 & \cos\theta \end{array} \right)
\left(\begin{array}{c} \tilde{S}_x \\ \tilde{S}_y \\
\tilde{S}_z \end{array} \right)
\end{equation}
in which $\theta=0$ for $h>1$,  $\theta=\arccos h$ for $0<h<1$ .
Then the Hamiltonian Eq. \eqref{h1} is converted into
\begin{eqnarray}
H &=& - \frac{1}{N}[\sin^2\theta \tilde{S_z}^2 + \frac{\cos^2\theta
+\gamma}{2}(\tilde{S}^2- \tilde{S_z}^2)]\nonumber\\
&+&
\frac{\sin2\theta}{N}(\tilde{S}^{+}\tilde{S_z}+\tilde{S}^{-}\tilde{S_z}+
h.c.)-\frac{\cos^2\theta -
\gamma}{4N}(\tilde{S}^{+2}+\tilde{S}^{-2})\nonumber\\&-& h_z
\cos\theta\tilde{S}_z
-\frac{h_z}{2}\sin\theta(\tilde{S}^{+}+\tilde{S}^{-}),
\end{eqnarray}
in which $\tilde{S}^{\pm}=\tilde{S}_x \pm i\tilde{S}_y$.

In order to obtain the geometric phase of ground state, we consider
the system has a rotation $\tilde{g}(\phi)$ around the new $z$
direction. The Hamiltonian becomes
\begin{equation}\label{h1phi}
H(\phi)=\tilde{g}(\phi)H\tilde{g}^{\dagger}(\phi).
\end{equation}
in which $\tilde{g}(\phi)=e^{i\phi\tilde{S}_z}$. Then we can use the
Holstein-Primakoff representation,
\begin{eqnarray}\label{hp}
\tilde{S}_z(\phi) &=& N/2 - a^{\dagger}a, \nonumber\\
\tilde{S}^{+}(\phi)&=&(N-a^{\dagger}a)^{1/2}a e^{i\phi},\nonumber\\
\tilde{S}^-(\phi)&=&a^{\dagger} e^{- i\phi}(N-a^{\dagger}a)^{1/2}
\end{eqnarray}
in which $a^{(\dagger)}$ is bosonic operator. Since the $z$ axis is
along the semiclassical magnetization, $a^{\dagger}a/N\ll 1$ is a
reasonable assumption under low-energy approximation, in which $N$
is large but finite. Keeping the terms of order $N, N^{1/2}, N^0$,
Eq. \eqref{h1phi} becomes
\begin{equation}
H(\phi)=Ne + \Delta a^{\dagger}a + \Gamma (a^{\dagger 2}e^{- 2i\phi}
+ a^2 e^{2i \phi}),
\end{equation}
in which
\begin{eqnarray}
e&=&- \frac{1}{4}(\sin^2\theta + 2h\cos\theta),\nonumber\\
\Delta&=&\sin^2\theta - \frac{\gamma+
\cos^2\theta}{2}+h\cos\theta\nonumber\\
\Gamma&=&\frac{\gamma-\cos^2\theta}{4}
\end{eqnarray}
Obviously the equation above can be diagonalized by the standard
Bogoliubov transformation,
\begin{equation}\label{bt}
b(\phi)=\cosh x a e^{i\phi} + \sinh x a^{\dagger}e^{-i\phi}
\end{equation}
Then the Hamiltonian is
\begin{equation}
H_{diag}(\phi)=Ne+\sigma + \Delta^D b^{\dagger}(\phi)b(\phi),
\end{equation}
in which,
\begin{eqnarray}
\sigma&=&\frac{\Delta}{2}(\sqrt{1-\epsilon^2}-1)\nonumber\\
\Delta^D&=&\Delta\sqrt{1 - \epsilon^2}\nonumber\\
\epsilon&=&\frac{2\Gamma}{\Delta}=\tanh2x=
\begin{cases}-\frac{1-\gamma}{2h-1-\gamma}, & h>1 \\
-\frac{h^2-\gamma}{2-h^2-\gamma}, & 0<h<1 \end{cases}
\end{eqnarray}
It should be careful about the physical interpretation of
$\Delta^D$, which may not describe true gap of the system
\cite{dusuel}.

Now it is time to determine the ground state $\ket{g(\phi)}$, which
can be obtained by applying the relation,
\begin{equation}
b(\phi)\ket{g(\phi)}=0
\end{equation}
Substituting Eq. \eqref{bt} into the equation above, one obtains the
ground state,
\begin{widetext}
\begin{eqnarray}
\ket{g(\phi)}=\frac{1}{C}\sum_{n=0}^{[N/2]}\sqrt{\frac{(2n-1)!!}{2n!!}}(-
\frac{e^{-i\phi}\sinh x}{e^{i\phi}\cosh x
})^{n-1}(-\sqrt{2}e^{-i\phi}\sinh x )\ket{2n},
\end{eqnarray}
\end{widetext}
in which $n!!=n(n-2)(n-4)\cdots$ and $n!!=1$ for $n\leq0$. $\ket{n}$
is the Fock state of bosonic operator $a^{(\dagger)}$ and the
normalized constant is
$C^2=\sum_{n=0}^{[N/2]}2\sinh^2x\frac{(2n-1)!!}{2n!!}\tanh^{2(n-1)}x$.
One should note that in order that the summation is convergent,
$|\tanh x| \leq 1$.

\begin{figure}
\begin{overpic}{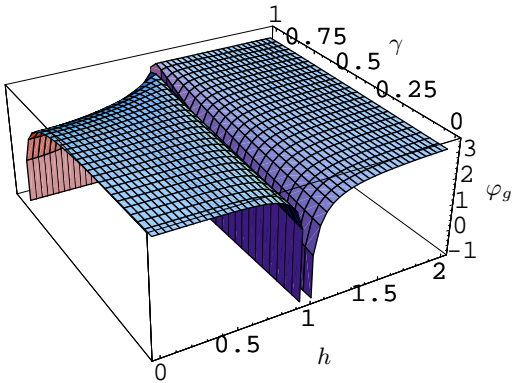}
\put(65, 5){$h$} \put(80, 70){$\gamma$} \put(100, 40){$\varphi_g$}
\end{overpic}
\caption{\label{g1} The geometric phase $\varphi_g$ [Arc] vs. the
anisotropic parameter $\gamma$ and $h$. For specification, we have
chosen the summation from 0 to 100 in the expression of $\varphi_g$
(Eq. \eqref{g}). The divergent character of $\varphi_g$ is clearly
displayed at $h\rightarrow1$ in this figure.}
\end{figure}
 The geometric phase of the ground state, accumulated by changing
$\phi$ from $0$ to $\pi$, is determined by
$\varphi_g=-i\int_0^{\pi}d\phi\bra{g(\phi)}\partial_{\phi}\ket{g(\phi)}$.
The direct calculation shows
\begin{equation}\label{g}
\varphi_g=\pi[1 -
\frac{\sum_{n=0}^{[N/2]}2n\frac{(2n-1)!!}{2n!!}\tanh^{2(n-1)}x}
{\sum_{n=0}^{[N/2]}\frac{(2n-1)!!}{2n!!}\tanh^{2(n-1)}x}]
\end{equation}
which have been plotted with  $\gamma$ and $h$ in Fig. \ref{g1}. It
is obvious that $\varphi_g$ is divergent at the point $h=1$ where
the LMG model has been proved to experience a second-order phase
transition, independent of the anisotropy $\gamma$\cite{botet}. This
divergency has never been explored in the previous studies
\cite{carollo, hamma, zhu, plastina} and shows distinguished
character from the XY  and Dicke models. In fact, since the
adiabatic condition has been destroyed when $h\rightarrow 1$ because
of the degeneracy  between the ground state and the  excited state
and the degeneracy enhances geometric phase \cite{berry}, this
phenomenon is natural. However, the essential reason is the
collective interaction in the LMG model, which is absent in the XY
model and makes the long-range correlations in this system.

\begin{figure}[t]
\begin{overpic}[width=4cm]{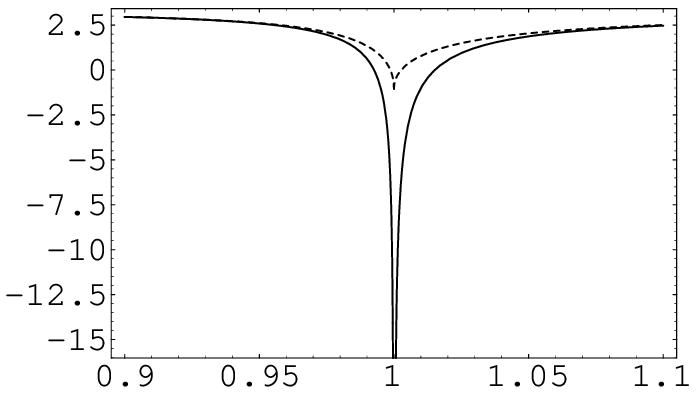}
\put(15,60){(a)}\put(50, -5){$h$} \put(-5, 35){$\varphi_g$}
\end{overpic}\vspace{1em}
\begin{overpic}[width=4cm]{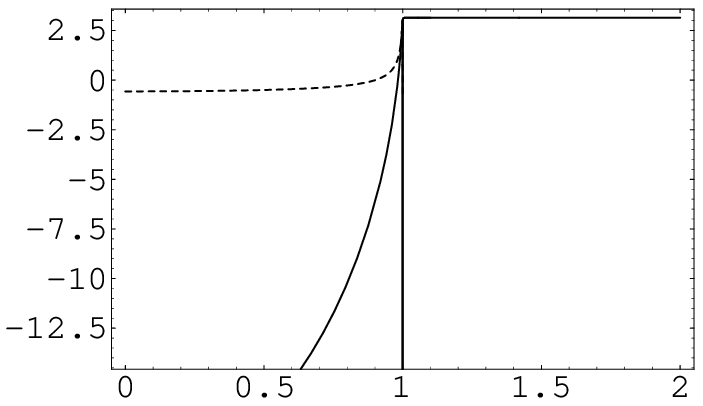}
\put(15,60){(b)}\put(50, -5){$h$} \put(-5, 35){$\varphi_g$}
\end{overpic}
\caption{\label{g11}$\varphi_g$ [Arc] vs. $h$ with different
particle number $N$. We have chosen $\gamma=0.5$ (a) and $\gamma=1$
(b) for this plot. The dashed and solid lines correspond
respectively to $N=4, 1000$.}
\end{figure}

The finite size effect is also examined in this case by choosing
different $N$, which can be shown in the Fig. \ref{g11}(a) and (b).
In this figure, we only draw for $N=4, 1000$ and for higher values
of $N$, the curves almost do not change. From the figures we note
that the divergency of $\varphi_g$ seems to be insensitive to $N$.
However, since phase transition only happens under thermodynamic
limit, this phenomenon attributes to the ground-state degeneracy at
the critical point.

\begin{figure}
\begin{overpic}[width=8cm]{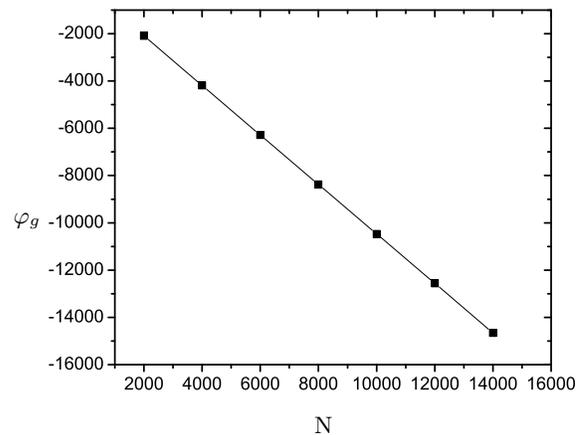}
\put(50, 0){N} \put(0, 35){$\varphi_g$} \end{overpic}
\caption{\label{scale} The scaling behavior of $\varphi_g$ [Arc] vs.
$N$ with $\gamma=0.5$ when $h\rightarrow1$. The slope of the line is
close to $-1$. }
\end{figure}
The scaling behavior of $\varphi_g$ plotted with $N$ has also been
explored in Fig. \ref{scale}. It is obvious that one can obtain
\begin{equation}
\varphi_g\approx- N.
\end{equation}
Furthermore the scaling is independent of $\gamma$, which means that
for different $\gamma$, the phase transitions belong to the same
university class. We also check the scaling behavior when $\gamma=1$
and find that this case has the same behavior as displayed in Fig.
\ref{scale}. It means that the phase transition  for $\gamma=1$ is
the same university class as that for $\gamma\neq1$. This phenomenon
is different from the XY model, in which the isotropic and
anisotropic interactions belong respectively to different university
classes \cite{zhu}, that comes from the collective interaction in
the LMG model.

\section{uniaxial model}
In this part, we will discuss the phase transition in a more complex
system, which is the generalization of the LMG model Eq. \eqref{h1}.
The Hamiltonian can be written as
\begin{equation}\label{h2}
H = - \frac{1}{N}S^2_x - h_x S_x -h_z S_z,
\end{equation}
with $h_z>0$. The phase diagram of this model is obviously dependent
on the both parameters $h_x, h_z$, and moreover a proper order
parameter characterizing the transition is difficult to build.
Recently the correspondence between this model and a two-level boson
problem introduced in nuclear physics has been constructed, which
permits one to get an order parameter \cite{vidal3}. The phase
transition is then clear in this model; a first-order transition
occurs at $h_x=0$ with $h_z<1$ and a second-order one occurs at
$h_x=0$ with $h_z=1$. For $h_z>1$ or $h_x\neq0$, no transition is
found.

Now we will try to determine the phase transition by calculating the
geometric phase of the ground state. Similar procedures as  in
previous section can be applied for this purpose. The first step is
to make the system have a rotation around the $z$ direction and then
the Hamiltonian \ref{h2} becomes
$H(\phi)=g(\phi)Hg^{\dagger}(\phi)$. Next step is to introduce the
Holstein-Primakoff transformation Eq. \eqref{hp}. One should note
that the approximation $a^{\dagger}a/N \ll 1$ is invalid in this
case since we cannot find the semiclassical magnetization. However a
simple canonical transformation can be used to resolve this problem,
$a^{(\dagger)}e^{(-)i\phi}=b^{(\dagger)}e^{(-)i\phi}+
\sqrt{N}\lambda$ with $|\lambda|<1$. This transformation provides a
macroscopic expectation value of $S_z$ which is order of $N$, and
then one has $b^{\dagger}b/N\ll1$. Under the limit that $N$ is large
but finite, it is enough to expand $H(\phi)$ to the order $N^0$.
After the exhausted calculation, $H(\phi)$ is written as,
\begin{eqnarray}\label{h2phi}
H(\phi)=&&e_0(\lambda)+
\Omega(\lambda)(be^{i\phi}+b^{\dagger}e^{-i\phi})+\nonumber\\
&&\Gamma(\lambda)(b^2e^{2i\phi}+b^{\dagger
2}e^{-2i\phi})+\Delta(\lambda) b^{\dagger}b,
\end{eqnarray}
in which,
\begin{eqnarray}
e_0(\lambda)&=& -N[\frac{h_z}{2}(1 -
2\lambda^2)+\lambda^2(1-\lambda^2)+h_x(1-\lambda^2)]\nonumber\\&-&
(\frac{1}{4}-\lambda^2)- h_x\frac{\lambda(2-\lambda^2)}{8(1 -
\lambda^2)^{3/2}}\nonumber\\
\Omega(\lambda)&=&\sqrt{N} [\lambda h_z - \frac{h_x(1-
2\lambda^2)}{2\sqrt{1-\lambda^2}}-\lambda(1-2\lambda^2)]\nonumber\\
\Gamma(\lambda)&=&-\frac{1-
5\lambda^2}{4}+h_x\frac{\lambda(2-\lambda^2)}{8(1-\lambda^2)^{3/2}}\nonumber\\
\Delta(\lambda)&=&h_z -
\frac{1-7\lambda^2}{2}+h_x\frac{\lambda(4-3\lambda^2)}{4(1-\lambda^2)^{3/2}}.
\end{eqnarray}
The crucial step is to choose $\lambda_0$ properly in order that the
linear term (the second term in Eq. \eqref{h2phi}) is vanishing. It
can be realized by solving the following equation,
\begin{equation}
\lambda_0 h_z - \frac{h_x(1-
2\lambda_0^2)}{2\sqrt{1-\lambda_0^2}}-\lambda_0(1-2\lambda_0^2)=0.
\end{equation}
In fact the equation above can be reduced into the biquadratic
equation $(h_z - y)^2(1-y^2)-h_x^2y^2=0$ with $y=1- 2\lambda_0^2$,
which can be solved numerically.

Substitute $\lambda_0$ into Eq. \eqref{h2phi} and then one get the
quadratic Hamiltonian,
\begin{equation}
H=e_0(\lambda_0)+\Gamma(\lambda_0)(b^2e^{2i\phi}+b^{\dagger
2}e^{-2i\phi}) +\Delta(\lambda_0)b^{\dagger}b,
\end{equation}
which obviously could be diagonalized by the standard Bogoliubov
transformation. Consequently the geometric phase of ground state can
be determined directly, which has the same form as Eq. \eqref{g},
but different definition of $x$, determined by the equation $\tanh
2x=\frac{2\Gamma(\lambda_0)}{\Delta(\lambda_0)}$.

\begin{figure}
\begin{overpic}{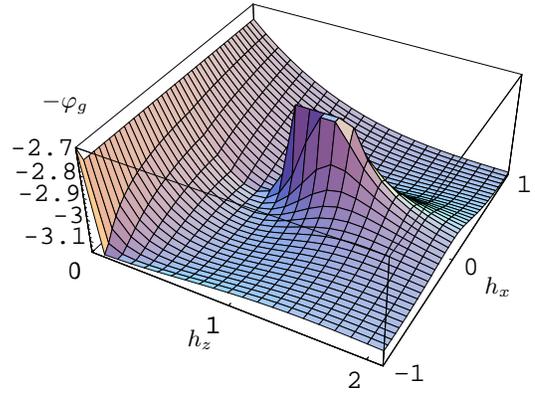}
\put(34, 10){$h_z$} \put(90, 20){$h_x$} \put(5, 55){ $-\varphi_g$}
\end{overpic}
\caption{\label{g2} The geometric phase $\varphi_g$ [Arc] vs. $h_x$
and $h_z$. We have chosen the summation from 0 to 100 in the
expression of $\varphi_g$ Eq. \eqref{g}, and for the convenience of
viewport, we have drawn for $-\varphi_g$ in this plot.}
\end{figure}
\begin{figure}[t]
\begin{overpic}[width=4cm]{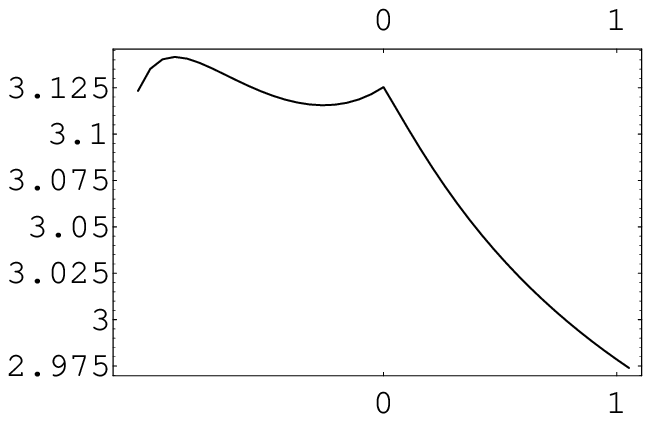}
\put(10,65){(a)}\put(50, -5){$h_x$} \put(-5, 35){$\varphi_g$}
\end{overpic}\vspace{1em}
\begin{overpic}[width=4cm]{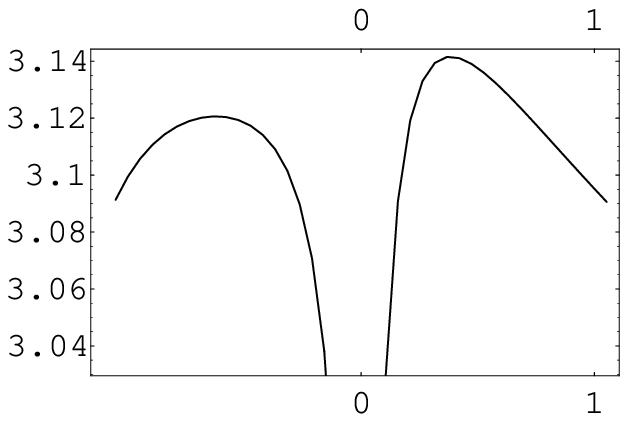}
\put(10,65){(b)}\put(50, -5){$h_x$} \put(-5, 35){$\varphi_g$}
\end{overpic}
\begin{overpic}[width=4cm]{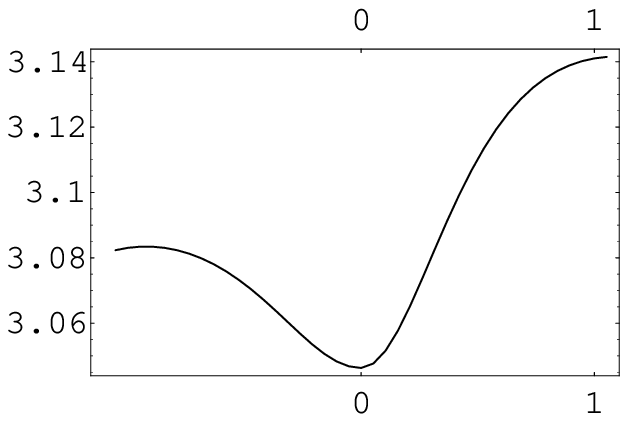}
\put(10,65){(c)}\put(50, -5){$h_x$} \put(-5, 35){$\varphi_g$}
\end{overpic}
\caption{\label{g22}The geometric phase $\varphi_g$ [Arc] vs. $h_x$
with different $h_z$. We have chosen $h_z=0.5$ (a),  $h_z=1$ (b) and
$h_z=2$ (c) for this plot. The figure (b) has been compressed
because of the divergency of the value $\varphi$.}
\end{figure}
A schematic demonstration of the geometric phase is presented in
Fig. \ref{g2}. It is obvious that there are two regions divided by
$h_z=1$, and the geometric phase is divergent at point $h_x=0,
h_z=1$. In the region $h_z<1$, geometric phase is non-analytical at
point $h_x=0$, which means the appearance of phase transition, and
in the other region, the geometric phase is the smooth function of
$h_z, h_x$ and no phase transition is found from the figure. This
phenomenon is consistent with the conclusion in Ref. \cite{vidal1},
but in our calculation we do not need to find a proper order
parameter to characterize the phase transition and the geometric
phase of ground state faithfully marks these transitions. A detailed
demonstration is also provided in Fig. \ref{g22}. From the figures,
one easily finds that the geometric phase $\varphi_g$ has a cusp at
$h_x=0$ for $h_z<1$ (see Fig. \ref{g22}(a)), which implies the
first-order phase transition. Furthermore $\varphi_g $ is divergent
at $h_x=0$ when $h_z=1$ (see Fig. \ref{g22}(b)), which is similar to
that in the standard LMG model (see Figs. \ref{g1}and \ref{g11}) and
means that there is a second-order transition. For $h_z>1$
$\varphi_g$ is the smooth function of $h_x, h_z$ and there is no
transition (see Fig. \ref{g22}(c)).

It is surprising that the geometric phase of ground state itself can
differentiate phase transition without need of introducing a proper
order parameter, which may be difficult to find. Thus it is a
natural speculation that the geometric phase of ground state can
serve as a universal order parameter. Further discussions will be
presented in the next part.

\section{conclusions and discussions}
The geometric phases of ground state in the LMG model and its
generalization have been discussed in this paper. Our calculations
show that the geometric phase  faithfully reflects the phase
transition in this model; when there is a second-order phase
transition, the geometric phase behaves divergent (see Figs.
\ref{g1} and \ref{g22}(b)). However when there is a first-order
transition, non-analyticity of geometric phase appears at the
critical point (see Fig. \ref{g22}(a)). These phenomena may come
from the degeneracy of the ground state in the system, which usually
induces the mixture of different phases. The different behaviors of
geometric phase closed to critical points could originate from the
different excitations (i.e. gapped or gapless). Furthermore we find
that the energy of ground state is not a good parameter of marking
the phase transition, because the energy of ground state is
degenerate independent of  the phase transition is second-order or
first-order. Recently Tian and Lin have shown that the continuous
quantum phase transition are actually caused by level crossing of
the low-lying excited states of the system \cite{tian}. This
conclusion also shows that the only energy of the ground state is
not suitable for the characterization of quantum phase transition.

This leads to a question what physical quantity is suitable for
characterizing the quantum phase transition. With respect of the
work Ref \cite{zhu} and our calculations, it seems to provide us an
hint that the geometric phase of ground state or its derivatives
could serve as an universal order parameter to characterize
different phase transitions. This speculation is natural since the
geometric phase faithfully measure the curvature of the Hilbert
space (or phase space for classical mechanics) and the broken of
symmetry of Hilbert space must be reflected in the geometric phase.

Another aspect of importance is the differences between our model
and the one-dimensional XY model. The crucial point is the
collective interaction in LMG model, which is absent in the XY
model. A main result of this interaction is that the LMG model
cannot be converted into the spinless fermion system. This fact
makes the conclusion different from that in Ref. \cite{hamma}, in
which the topological behavior of Berry phase can detect the
critical point. However, in this model, the geometric phase is
divergent when there is second-order transition and the detection
maybe invalid.

In conclusion the relation between the geometric phase of ground
state and the phase transition in LMG model has been constructed in
this paper. Different from the results in the one-dimensional XY
model, the singularity of the geometric phase itself can serve as
the signature of critical points. Moreover we discuss the
possibility of the geometric phase of ground state or its
derivatives serving as the universal order parameter to characterize
the quantum phase transitions.

This work was supported by NSF of China under grants 10305002 and
60578014.

\end{document}